\documentclass[conference]{IEEEtran}

\usepackage{amsmath}
\usepackage{url}
\usepackage{balance}

\usepackage{lambda}
\renewcommand{\NOTE}[1]{}

\usepackage{newtxtext,newtxmath}
\usepackage{MnSymbol}
\usepackage{mathtools}

\def\defmacro#1{\expandafter\def\csname#1\endcsname}
\def\newdef#1#2{\expandafter\ifx\csname#1\endcsname\relax
    \defmacro{#1}{#2}%
    \else\message{Command "#1" already defined}\fi}
\newif\ifmore
\def\mapdef#1#2(#3){\def\args{#1:#2:#3,\end}%
    \moretrue
    \loop\expandafter\nextarg\args
         \ifmore\repeat}
\def\nextarg#1:#2:#3,#4\end{\def\next{#4}%
    \ifx\next\empty\morefalse
        \else\def\args{#1:#2:#4\end}\fi
    \newdef{#2#3}{#1{#3}}}

\newcommand{\tyFmt}[1]{\textit{#1}}
\mapdef\tyFmt{}(Company,Quarter,Int,String,Unit,Name,LastName,Profit,Year)
\newcommand{\Val}{\textit{V}}
\newcommand{\Attr}{\textit{A}}
\newcommand{\Record}{\textit{R}}
\newcommand{\rec}{\textit{r}}
\newcommand{\AVal}{\ensuremath{\bar{V}}}
\newcommand{\RVal}{\ensuremath{\hat{V}}}
\newcommand{\DType}{\ensuremath{\Delta}}
\newcommand{\dty}{\ensuremath{\delta}}
\newcommand{\cty}{\ensuremath{\gamma}}
\newcommand{\rty}{\ensuremath{\rho}}
\newcommand{\tty}{\ensuremath{\tau}}
\newcommand{\Data}{\textit{D}}
\newcommand{\Table}{\textit{T}}

\newcommand{\valFmt}[1]{\textrm{#1}}
\mapdef\valFmt{}(Anne,Bob,Carl,Joe,Smith,Jones)

\newcommand{\QA}{\valFmt{Q1}}
\newcommand{\QB}{\valFmt{Q2}}
\newcommand{\QC}{\valFmt{Q3}}
\mapdef\valFmt{co}(A,B,C)

\newcommand{\refined}[2]{\ensuremath{#1\mathop\leftY#2}}
\newcommand{\rv}{\ensuremath{\hat{v}}}

\newcommand{\attr}[2]{\ensuremath{#2\mathord=#1}}
\newcommand{\aval}[2]{\ensuremath{#1^{#2}}}
\newcommand{\av}{\ensuremath{\bar{v}}}

\newcommand{\sequ}[1]{\ensuremath{\langle#1\rangle}}
\newcommand{\ty}[2]{\ensuremath{#1\sequ{#2}}}

\newcommand{\lkup}[2]{\ensuremath{#1\mathord\uparrow#2}}
\newcommand{\mkTabSym}{\ensuremath{\boxplus}}
\newcommand{\mkTab}[3]{\ensuremath{\prescript{#1}{#2}{\boxplus}#3}}

\newcommand{\reverse}[1]{\ensuremath{\textit{rev}(#1)}}
\newcommand{\filter}[2]{\ensuremath{\sigma_{#1}(#2)}}
\newcommand{\refine}[2]{\ensuremath{#1\otimes#2}}

\newcommand{\records}[1]{\ensuremath{\lfloor#1\rfloor}}

\newcommand{\SUMSym}{\textsc{sum}}

\newcommand{\MAX}[1]{\ensuremath{\textsc{max}(#1)}}

\begin{document}


\title{Typed Table Transformations}

\author{\IEEEauthorblockN{Martin Erwig}
\IEEEauthorblockA{%
\textit{Oregon State University}\\
erwig@oregonstate.edu}
}

\maketitle

\thispagestyle{plain}
\pagestyle{plain}

\begin{abstract}
Spreadsheet tables are often labeled, and these labels effectively constitute
types for the data in the table. In such cases tables can be considered to be
built from typed data where the placement of values within the table is
controlled by the types used for rows and columns.
We present a new approach to the transformations of spreadsheet tables that is
based on transformations of row and column types.
We illustrate the basic idea of type-based table construction and
transformation and lay out a series of research questions that should be
addressed in future work.
\end{abstract}


\section{Introduction}
\label{sec:intro}

Spreadsheets present data and computation with those data in tabular form.
As reported by Harris and Gulwani \cite{HG11pldi}, Excel users often face the
problem of transforming tables.
Consider, for example, the table in Figure \ref{fig:example}(a), which is
adapted from \cite{HG11pldi} and based on the actual transformation needs by an
Excel
user\footnote{\url{http://www.excelforum.com/excel-programming-vba-macros/698490-using-a-macro-to-extract-and-rearrange-data.html}}.
In our version the table shows the earnings of three companies for three
different quarters. Suppose we want to transform this table into a list that
shows individual earnings in separate rows for each company and quarter,
ignoring empty cells for non-existing data. The result should look like the
table shown in Figure \ref{fig:example}(b).

Harris and Gulwani describe in their paper an algorithm that can infer
transformations of tables such as from (a) to (b) from input/output examples.
In this example, as in many others, we can observe that the rows and columns of
tables contain labels that explain the (purpose of the) data in the table.
These labels can be interpreted as type information for the data in the table
\cite{EB02padl,AEA07hcc}.
In this paper we present an approach that exploits this fact and lets users
describe table transformations based on transformation of their row and column
types.

Based on the concepts of row and column types and typed tables, we describe in
Section \ref{sec:tt} how tables can be systematically constructed from
attributed data driven by types.
In Section \ref{sec:typeop} we then show how table transformations can be
expressed through transformations and operations on their types.
We discuss some related work in Section \ref{sec:rw} and present conclusions in
Section \ref{sec:concl} where we also lay out a plan for future work. Since
this is a short paper reporting on work in progress, the focus is on explaining
the major ideas and identifying research questions to be addressed in the
future.

\begin{figure}[tb]
{\small
\begin{tabular}[b]{@{}l}
\begin{tabular}[b]{|c|c|c|c|} \hline
      & \QA & \QB & \QC \\ \hline
\coA &  3.5 & 2.9 & 4.0 \\ \hline
\coB  & 3.2 &     & 4.3 \\ \hline
\coC &      & 4.9 &     \\ \hline
\end{tabular}
\\[1ex]
(a) Original Table
\end{tabular}
\hfill
\begin{tabular}[b]{l}
\begin{tabular}[b]{|c|c|c|} \hline
\coA & \QA & 3.5 \\ \hline
\coA & \QB & 2.9 \\ \hline
\coA & \QC & 4.0 \\ \hline
\coB & \QA & 3.2 \\ \hline
\coB & \QC & 4.3 \\ \hline
\coC & \QB & 4.9 \\ \hline
\end{tabular}
\\[1ex]
(b) Linearized
\end{tabular}
\hfill
\begin{tabular}[b]{l@{}}
\begin{tabular}[b]{|c|c|} \hline
\coA & 4.0 \\ \hline
\coB & 4.3 \\ \hline
\coC & 4.9 \\ \hline
\end{tabular}
\\[1ex]
(c) Aggregated
\end{tabular}
}
%
\caption{(a) A table showing the earnings of three companies in three different
quarters. (b) The same data in linearized form. (c) The data for each company
aggregated over all quarters.}
\label{fig:example}
\end{figure}

\section{Typed Tables}
\label{sec:tt}

The data items in a two-dimensional table are uniquely identified by their row
and column positions. When the rows and columns are labeled, these labels can
serve as names for the table positions, which provides a more high-level,
domain-centered way for talking about data placement in tables.
For example, to find the value 2.9 we can look into the second row and third
column of the table \ref{fig:example}(a), or we can look up the value for
company \coA\ and quarter \QB.
In the following we formalize this idea.

\subsection{Values and Types}

In the context of spreadsheets, values ($v\in\Val$) include simple data types
such as numbers, dates, or strings. We use the metavariable $n$ to range over
those values (typically strings) that are used as type names.

A \emph{domain type} ($\dty\in\DType$) consists of a name and a finite set of
values.
A domain type is different from predefined types such as \Int\ or \String\
ordinarily found in programming languages: It is defined by a user and is used
to indicate the nature, origin, or purpose of other values.
A simple example of a domain type is \ty{\Company}{\coA, \coB, \coC} where
\Company\ is the name of the type and \coA, \coB, and \coC\ are its values.
Domain types such as \Company\ that contain only plain values are called
\emph{plain}. Otherwise, they are called \emph{refined}. We will see
types with refined values later in Section \ref{sec:refine}.

An \emph{attribute} is a value, such as \Joe, associated with a type name, such
\Name, and is written as \attr{\Joe}{\Name}. Note that attributes can be formed
arbitrarily; in particular, the value does not have to be an element of the
associated type.
A set of attributes is called a \emph{record}, and a value with an associated
record is called an \emph{attributed value}.

%
%

Finally, a \emph{table type} ($\tty\in\dty\times\dty$) consists of a pair of
domain types, the first representing the column type and the second
representing the row type, and a table $t\in\Table$ is a mapping from
addresses, represented as pairs of natural numbers, to values.
%
%
The syntax of values and types is summarized in Figure \ref{fig:syntax}. The
attentive reader will notice that we do not consider formulas in this model.

\begin{figure}[tb]
$\begin{array}{l@{\qquad}r@{\ \in\ }l@{\ }c@{\ }l}
\textrm{Values \& Type Names} & v,n & \Val \\
\textrm{Refined Values}       & \rv & \RVal &::=& v\ |\ \refined{v}{\dty}
\\[2ex]
\textrm{Domain Types} & \dty,\cty,\rty & \DType &::=& \ty{n}{\rv^*} \\
\textrm{Table Types}  & \tty & \multicolumn{3}{@{}l}{\dty\times\dty}
\\[2ex]
\textrm{Attributes}        & a   & \Attr   &::=& \attr{v}{n} \\
\textrm{Records}           & \rec& \Record &::=& \set{a^*} \\
\textrm{Attributed Values} & \av & \AVal   &::=& \aval{v}{\rec} \\
\textrm{Attributed Data}   & d   & \Data  &=& \smash{2^{\AVal}} \\
\textrm{Tables}            & t   & \Table &=& \mathbb{N}\times\mathbb{N}\to\Val
\end{array}$
\renewcommand{\arraystretch}{1.0}
\caption{Syntax of values and types}
\label{fig:syntax}
\end{figure}

\subsection{Tables}

Our approach to table transformations is based on the premise that tables are
the result of the systematic presentation of attributed values. Specifically,
the construction of tables is driven by types that are associated with their
rows and columns.
For example, the table in Figure \ref{fig:example}(a) is the result of
creating a table with column type \ty{\Quarter}{\QA,\QB,\QC} and row type
\ty{\Company}{\coA,\coB,\coC} from a set of attributed values such as the
following:
\[
\set{\aval{2.9}{\set{\attr{\coA}{\Company},\attr{\QB}{\Quarter},\ldots}},
     \aval{3.2}{\set{\attr{\coB}{\Company},\attr{\QA}{\Quarter},\ldots}},
     \ldots}
\]
For example, the first attributed value in this set is 2.9, which has (at
least) the two attribute values \coA\ of type \Company\ and \QB\ of type
\Quarter; it may have further attributes, but these two are relevant for the
proper placement of the value in the $\Quarter\times\Company$ table, which
works by looking up the position of the two attributes in their respective
types. Since \QB\ is the second element of the type \Quarter\ and \coA\ is the
first element of the type \Company, the value 2.9 is placed in column 2 and row
1 of the core table.\footnote{The table row and column headers that are given
by the values of the corresponding row and value types are later added to the
core table and are ignored in this calculation.}
In the same fashion all the other values will be positioned based on their
attributes.

The locations of empty cells in the table correspond to attribute combinations
for \Quarter\ and \Company\ attributes that do not occur with values in the
data set.

This approach to the construction of tables can be formalized by a
function
%
$\mkTabSym:\DType\times\DType\times\Data\to\Table$.
The definition employs the auxiliary lookup function \lkup{\dty}{r}, which
searches for an attribute \attr{v}{n} in the record $r$ given domain type
$\dty=\ty{n}{\rv_1,\ldots,\rv_k}$ and, if found, determines $v$'s position
among \dty's values $\rv_1,\ldots,\rv_k$, which then provides the row or column
for $v$ in the constructed table. The definition for $\uparrow$ is obvious if
all the $\rv_i$ are plain values.
The case for types with refined values is more involved and will be discussed
later.
The definition of \mkTabSym\ is now straightforward. Note that we generally
employ the metavariable \cty\ for column types and \rty\ for row types.
\[
\mkTab{\cty}{\rty}{D} =
  \set{((x,y),v)\ |\ \aval{v}{r}\in D \wedge \lkup{r}{\cty}=x
                                      \wedge \lkup{r}{\rty}=y}
\]
Note that the function \mkTabSym\ only builds the core part of the table
consisting of the data values. In addition, we need to add the column and row
headers. Since the corresponding definitions are not very interesting, we omit
them here for brevity.

Assuming that the data source of attributed values is $D$, we can now
construct the table shown in Figure \ref{fig:example}(a) with the following
expression.
\[
t=\mkTab{\Quarter}{\Company}{D}
\]
From the definition of \mkTabSym\ we can immediately infer the following
properties of table construction.

First, as already mentioned, values whose attributes do not contain values of
both row and column type will be not placed, that is, data with insufficient
attributes are simply ignored.

Second, when data items have the same attribute values for the row and column
types, the set definition does not produce a function and is effectively
undefined in terms of its result type (which is
$\Table=\mathbb{N}\times\mathbb{N}\to\Val$). In this case the table
construction simply fails, since  different values would be mapped to the same
locations, resulting in an ambiguity.
This interpretation is probably too strict, since many application scenarios
could benefit from constructing tables with underspecified type information.
These cases can be handled in two different ways: (1) We can preserve
multiple values by mapping a row/column combination not just to a single cell
but to a group of cells. This complicates the computation of cell locations,
but is otherwise not difficult to achieve in principle. (2) We can apply
aggregating functions such as \SUMSym\ to aggregate a set of values into one
value.

Third, it is easy to see that tables built with \mkTabSym\ can be transposed by
simply exchanging the column and row type, that is, we know the following
identity holds.
\[
(\mkTab{\cty}{\rty}{D})^{\textrm{T}} =\ \mkTab{\rty}{\cty}{D}
\]

\section{Table Transformations Through Type Operators}
\label{sec:typeop}

Since the structure of tables built with \mkTabSym\ depends on the row and
column types, it is not surprising that changes to these types result in
corresponding changes for the constructed tables.
In this section we present different kinds of type transformations and discuss
how they give rise to corresponding table transformations.

\subsection{Transforming of Row and Column Types}

Consider again the table shown in Figure \ref{fig:example}(a) and defined
as $t$ in the previous section.
Suppose now that we want this table to show the data only for companies \coA\
and \coC\ and also only for the first and third quarter.
We can apply the selection criteria to the row and column types using a
selection operation $\sigma_P(\dty)$ to filter out all elements from type
\dty\ that do not satisfy the predicate $P$.
By using filtered domain types in the \mkTabSym\ operation we can build the
correspondingly amended table.\footnote{We use the Purescript notation for
partial function application in which a binary operation applied to one of its
arguments denotes a function of its other remaining argument, for example,
$\_\mathord\neq x \equiv \lambda y.y\mathord\neq x$.}
\[
\mkTab{\filter{\_\neq\QB}{\Quarter}}
      {\filter{\_\neq\coB}{\Company}}{D}
\]
Especially for bigger tables and with more complicated selection criteria,
reconstructing the table in this way is probably faster and less error-prone
than directly editing the table in Excel by repeatedly cutting and pasting rows
and columns.

Note that we can achieve the same effect by filtering the data source $D$
directly with a conjunction of the two predicates.
\[
\mkTab{\Quarter}{\Company}
      {(\filter{\Company\neq\coB\wedge\Quarter\neq\QB}{D})}
\]
In general, when $\cty=\ty{n}{\ldots}$ and $\rty=\ty{m}{\ldots}$, we can
observe the following equivalence between type and data selection.
\[
\mkTab{\filter{Q}{\cty}}{\filter{P}{\rty}}{D}\ =\
\mkTab{\cty}{\rty}{(\filter{P(n)\wedge Q(m)}{D})}
\]
Why then should we bother about the manipulation of row and column types?
Having selection available for table types has several potential advantages.
First, the row and column type selections show more directly the effect of the
selection on the structure and shape of the table than does the selection of
the data.
Second, and more importantly, some effects on table restructuring cannot be
easily achieved by manipulating the data. Consider, for example, the task of
reordering the rows of table $t$. This can be accomplished through the
following expression.
\[
\mkTab{\Quarter}{{\reverse{\Company}}}{D}
\]
However, it is not clear how to achieve this effect through a transformation of
$D$.

We can envision a number of other operations on row and column types that can
be put to use in the manipulating tables. For example, dually to filtering
types we can also extend types by new values, which amounts to growing tables
by row or columns. Of course, this produces results only if the underlying data
also contain correspondingly attributed values.

\NOTE{What about concatenating ($\approx$ union) of types? Does
\mkTab{{\Quarter\cup\Company}}{\Company}{D} make any sense?
How about:
\mkTab{\Quarter}{\Quarter}{D}? This yields a completely empty table, since each quarter is an attribute to more than one value. In contrast,
\mkTab{\Company}{\Company}{D} yields a table with exactly one entry, 4.9, in the lower right corner.
}

\subsection{Type Refinement}
\label{sec:refine}

A close look at the definition of \dty\ in Figure \ref{fig:syntax} reveals that
a type is essentially a tree structure with a type name as its root, values as
the root's children, and potentially other types as subtrees of children that
are refined.
For plain types these trees are rather trivial and have height 2, and
consequently, table construction with plain types covers only rather simple,
albeit useful, application scenarios.

Refined types correspond to trees with a more complex structure, and so more
sophisticated table constructions and transformations can be achieved with type
operations that create or modify types containing refined values.

The first such operation is type refinement \refine{\dty'}{\dty} that refines a
domain type $\dty'$ by another domain type \dty\, which means to attach the
whole type \dty\ to every leaf value in $\dty'$. The formal definition is as
follows.
\[
\begin{array}{r@{\ }l}
\refine{\ty{n}{\rv_1,\ldots,\rv_k}}{\dty} &=
        \ty{n}{\refine{\rv_1}{\dty},\ldots,\refine{\rv_k}{\dty}} \\
&\hphantom{=\ }\textrm{where}\
\begin{array}[t]{@{}r@{\ }l}
\refine{(\refined{v}{\dty'})}{\dty} &= \refined{v}{(\refine{\dty'}{\dty})} \\
                   \refine{v}{\dty} &= \refined{v}{\dty}
\end{array}
\end{array}
\]
The notion of type membership gets a bit more interesting for types with
refined values, since values occur at multiple levels. In fact, each path from
the root to a leaf represents a record of $(n,v)$ pairs. In the following we
define a function \records\_\ that computes for each type the sequence of
records it represents. In the definition we use $\cdot$ to denote the
concatenation of sequences and $\cup$ for computing the union of two records.
\[
\begin{array}{r@{\ }l}
\records{\ty{n}{\,}}       &= \sequ{\,} \\
\records{\ty{n}{v,\rv^*}}  &= \sequ{\set{\attr{v}{n}}}\cdot\records{\ty{n}{\rv^*}} \\
\records{\ty{n}{\refined{v}{\dty},\rv^*}}
      &= \sequ{\set{\attr{v}{n}}\cup r\ |\ r\in\records{\dty}}
         \cdot\records{\ty{n}{\rv^*}}
\end{array}
\]
For a type without any refined values, the records contain only single
attributes.
\[
{\footnotesize
\records{\ty{\Company}{\coA,\coB}} = \sequ{\set{\attr{\coA}{\Company}}, \set{\attr{\coB}{\Company}}}
}
\]
The following example illustrates how refined values expand records into
multiple attributes.
\[
{\footnotesize
\begin{array}{@{}l@{}l}
\records{\ty{\Company}{\refined{\coA}{\ty{\Quarter}{\QA,\QB}},\coB}} =\ & \langle\set{\attr{\coA}{\Company},\attr{\QA}{\Quarter}}, \\
&\hphantom\langle
      \set{\attr{\coA}{\Company},\attr{\QB}{\Quarter}}, \\
&\hphantom\langle
      \set{\attr{\coB}{\Company}}\rangle
\end{array}
}
\]
With this extended semantics of types, the placement of attributed values in
tables has to be adapted.
Specifically, instead of locating the position of a value, we now locate the
position of a record that is subsumed by the record of the attributed value to
be placed. We write $s_i$ for selecting the $i$th element from the sequence
$s$. As a special case, we consider the type \Unit, which consists of just one
single element and which is used to specify untyped single rows or columns of
tables: The position of any value is always 1 with respect to the type \Unit.
\[
\lkup{r}{\dty} \Begindef
  1  & \If \dty=\Unit \\
  i\ \textrm{such that}\ r\supseteq\records{\dty}_i & \Otherwise
\Enddef
\]
If domain types do not contain duplicates, there will be at most one record
that $r$ can subsume, and $i$ is unambiguously defined.
With this amended definition of value lookup, the previous definition for
\mkTabSym\ works now for types with and without refined values.

We can now try to construct the table shown in Figure \ref{fig:example}(b) with
the following expression.
\[
\mkTab{\Unit}{\refine\Company\Quarter}{D}
\]
However, the result of this expression is not quite the table shown in Figure
\ref{fig:example}(b), since it still contains empty cells just as the original
table in Figure \ref{fig:example}(a) does. We can remove those by using a table
filtering function that eliminates empty rows (and columns) from a table (using
a predicate \textit{empty} that is true for an empty data row or column).
\[
\filter{\neg \textit{empty}}{\mkTab{\Unit}{\refine\Company\Quarter}{D}}
\]
If this behavior is needed frequently or maybe even the expected default, we
can easily define a corresponding version of \mkTabSym\ that applies the filter
by default.

Note that we can put such filter functionality to a much wider use. For
example, we can produce a list of all companies and quarters for which
\emph{no} data is available.
\[
\filter{\textit{empty}}{\mkTab{\Unit}{\refine\Company\Quarter}{D}}
\]
Since the table construction is driven exclusively by the row and column types,
interesting table variations can be achieved by simple type transformations.
%
%
Suppose, for example, that we want to show quarters first and companies nested
inside of quarters. We can accomplish this transformation by changing the order
of type refinement.
\[
\mkTab{\Unit}{\refine\Quarter\Company}{D}
\]
This flexibility of type refinement is due to its \emph{not} being a
commutative operations, that is:
\[
\dty'\neq\dty\implies\refine{\dty'}{\dty}\neq\refine{\dty}{\dty'}
\]
Type refinement is very similar to cartesian product. The difference is the
tree-shape representation in which the refining type is attached to each value
of the type that is being refined. This provides additional flexibility for
further type operations. For example, we can define operations for selectively
removing or adding refinements for individual values. In this sense, types with
refined values bear some similarity to dependent sum types in type theory.

\subsection{Other Type Operations}

We can envision several other interesting type operations that can be exploited
for new forms of table constructions or transformations.
One such operation is \emph{type coarsening}, which is the inverse of
refinement and extracts a type from refined values. This operation can be used
to transform nested list structures into tables.
We could employ coarsening to transform tables of the kind shown in Figure
\ref{fig:example}(b) into tables of the from shown in Figure
\ref{fig:example}(a). In other words, inverse type transformations define
inverse table transformations.
%

More interestingly, though, coarsening can be used in connection with
aggregating functions to produce summary tables. For example, if we coarsen the
type \refine\Company\Quarter\ back to \Company, we have multiple values in the
data source matching each company, which makes the table construction
ambiguous. By aggregating a collection of values into a single value with a
binary function, we can get back a well-defined behavior. For example, the
following construction produces the table shown in Figure \ref{fig:example}(c).
\[
\mkTab{\Unit}{\Company}{\MAX{D}}
\]
Note that the binary function does \emph{not} have to be commutative or
associative, since the values to be aggregated are ordered. The function
doesn't need an identity element either, since empty cells can simply be kept
and don't need to be aggregated.


\section{Related Work}
\label{sec:rw}

This work is inspired by the work of Harris and Gulwani \cite{HG11pldi} in
which they present a programming-by-example approach to infer table
transformations from input/output example tables.
They describe an algorithm ProgFromEx that represents inferred table
transformations in a language called TableProg.
We haven't performed a detail comparison of the two approaches yet, but the
obvious advantages of ProgFromEx are that it needs no programming at all by end
users and can probably deal easier with some more complicated, unstructured
cases.
The main advantage of our approach is that the transformation results are
better predictable because transformations are based on a simple and clear
semantics instead of a complicated inference algorithm. Moreover, our table
transformations are highly reusable and composable, which promises better
scalability.
However, a detailed comparison of the two approaches is subject of future work.

Pivot tables as found, for example, in Excel can also be used to transform
tabular data. While the input table basically represents an attribute data set
(like a relation in a relational database), constructing Pivot tables is
primarily an interactive process and not based on explicitly applying table
transformation operations. Since Excel Pivot tables don't have any associated
notion of types, they obviously cannot provide operations for type
transformations.
However, given the strong similarities, typed table transformation could be
used as an underlying formal model for \emph{typed} Pivot tables.

The table transformations we have considered here are ad hoc in the sense that
they are based on arbitrary type transformations.
However, a large class of systematic table transformations are the result of
controlled evolution of table data. Several approaches have been proposed to
capture such evolution-based table transformations
\cite{EACK05icse,AE06icse,EE05ase,CEMS16ase}.
An interesting topic for future work is to identify type transformations that
can represent evolution-based table transformations.


\section{Conclusion and Future Work}
\label{sec:concl}

We have demonstrated that table types and transformations can be an effective
basis for the systematic construction and manipulation of spreadsheet tables.
%
%
However, the presented approach is incomplete and requires several additional
components to become a versatile and widely applicable table manipulation tool.
In particular, in future work we plan to address the following questions.

\begin{itemize}

\item How do we obtain labeled data sources in the first place?
We can have users mark areas of spreadsheets as data sources. By using an
interactive tool or label inference \cite{AE04hcc,AE07jvlc}, values can be
turned into attributed values.

\item How do we transform tables with formulas?
To be able to place formulas arbitrarily, we have to base cell references on
labels instead of addresses, just as we did in \cite{LEE12jvlc} or
\cite{CDEFG18gpce}. For generating concrete references when constructing
tables, we also have to distinguish between aggregating and iterating
operations \cite{EACK06jfp}.

\item We need a comprehensive language definition that includes operations for
turning tables into attributed data as well as combining different tables.
\end{itemize}
Tables are the fundamental data structure of spreadsheets. Investigating the
properties of tables as well as their transformations should thus be a priority
of spreadsheet research.

\section*{Acknowledgments}

This work is partially supported by the National Science Foundation under the
grant CCF-1717300.

\balance

\bibliographystyle{IEEETran}
\bibliography{me,sprsh}

\end{document}